\documentclass{article}

\usepackage{PRIMEarxiv}

\usepackage[utf8]{inputenc} 
\usepackage[T1]{fontenc}    
\usepackage{hyperref}       
\usepackage{url}            
\usepackage{booktabs}       
\usepackage{amsfonts}       
\usepackage{nicefrac}       
\usepackage{microtype}      
\usepackage{lipsum}
\usepackage{fancyhdr}       
\usepackage{graphicx}       
\usepackage{todonotes}

\usepackage{amsmath}
\DeclareMathOperator*{\argmax}{argmax}

\newcommand{\norm}[1]{\left\lVert#1\right\rVert}
\newcommand{\yobs}{\boldsymbol{y}_{\mathrm{obs}}}
\newcommand{\thetaMAP}{\boldsymbol{\theta}_{\mathrm{MAP}}}

\usepackage{changepage}
\usepackage{enumitem}
\usepackage{xcolor}
\usepackage{needspace}

\usepackage[style=numeric, sorting=none, giveninits=true, isbn=false,eprint=false, minnames=15, maxnames=20, backend=biber]{biblatex}
\bibliography{main.bib}  
\AtEveryBibitem{\clearfield{note}}
\AtEveryBibitem{\clearlist{language}}
\AtEveryBibitem{\clearfield{urlyear}}
\AtEveryBibitem{\ifentrytype{misc}{%
  }{%
    \clearfield{url}%
  }%
}

\definecolor{cblue}{rgb}{0.050383, 0.029803, 0.527975}
\definecolor{cred}{rgb}{0.881443, 0.392529, 0.383229}
\definecolor{cyellow}{rgb}{0.988648, 0.809579, 0.145357}

\hypersetup{
     linkbordercolor=cblue,
     citebordercolor=cred,   
     urlbordercolor=cyellow
}

\usepackage{algorithm2e}
\DontPrintSemicolon
\SetKwRepeat{Do}{do}{while}
\SetKwData{Input}{Input}
\usepackage[]{newfloat}
\DeclareFloatingEnvironment[fileext=frm,placement={!ht},name=Algorithm, listname=List of Algortihms]{algo}
\usepackage{caption}
\title{Solving Bayesian Inverse Problems With Expensive Likelihoods Using Constrained Gaussian Processes and Active Learning
}

\author{
  Maximilian Dinkel, Carolin M. Geitner, Gil Robalo Rei, Jonas Nitzler\\
  Institute for Computational Mechanics \\
  Technical University of Munich \\
  Munich \\
  \texttt{\{maximilian.dinkel, carolin.geitner, gil.rei, jonas.nitzler\}@tum.de} \\
   \And
  Wolfgang A. Wall \\
  Institute for Computational Mechanics, Munich Data Science Institute\\
  Technical University of Munich \\
  Munich \\
  \texttt{\{wolfgang.a.wall\}@tum.de} \\
}

\begin{document}
\maketitle

\begin{abstract}
Solving inverse problems using Bayesian methods can become prohibitively expensive when likelihood evaluations involve complex and large scale numerical models. A common approach to circumvent this issue is to approximate the forward model or the likelihood function with a surrogate model. But also there, due to limited computational resources, only a few training points are available in many practically relevant cases. Thus, it can be advantageous to model the additional uncertainties of the surrogate in order to incorporate the epistemic uncertainty due to limited data. In this paper, we develop a novel approach to approximate the log likelihood by a constrained Gaussian process based on prior knowledge about its boundedness. This improves the accuracy of the surrogate approximation without increasing the number of training samples. Additionally, we introduce a formulation to integrate the epistemic uncertainty due to limited training points into the posterior density approximation. This is combined with a state of the art active learning strategy for selecting training points, which allows to approximate posterior densities in higher dimensions very efficiently. We demonstrate the fast convergence of our approach for a benchmark problem and infer a random field that is discretized by 30 parameters using only about 1000 model evaluations.
In a practically relevant example, the parameters of a reduced lung model are calibrated based on flow observations over time and voltage measurements from a coupled electrical impedance tomography simulation. \end{abstract}

\keywords{Bayesian Inverse Problem \and Gaussian Process \and Active Learning}

\section{Introduction}

Computational modeling of physical phenomena has gained huge importance in all fields of engineering and applied science. However, most models rely on parameters that can not be measured directly and have to be calibrated. Some selected recent applications from our own group range from photoacoustic image reconstruction \cite{schoeder_photoacoustic_2017}, calibration of 
multiphase porous media model parameters to predict tumor growth \cite{hervas-raluy_tumour_2023}, the determination of diffusion parameters for laser powder bed fusion models \cite{nitzler_novel_2021} to the analysis of material parameters in
coupled multi-physics biofilm models \cite{willmann_inverse_2022}. A common approach is to solve this problem by formulating it as a least-square minimization problem with a regularization term \cite{tikhonov_numerical_1995, kaipio_statistical_2005}. However, this leads to ill-posed problems and a.o., the resulting point estimates do not capture measurement errors and can be misleading. By formulating the calibration task as a Bayesian inverse problem, the uncertainty of the measurements can be incorporated, and additionally, prior knowledge can be integrated. Rather than a point estimate, the Bayesian solution provides a probability distribution for the uncertain parameters given the noisy measurements. For practically relevant problems, the inference task can not be solved analytically. Nevertheless, the distribution can be approximated using sampling \cite{metropolis_equation_2004, neal_improved_1994, hoffman_no-u-turn_2014}, particle \cite{chopin_sequential_2002, chopin_introduction_2020, del_moral_sequential_2006, doucet_sequential_2001}, or variational methods \cite{blei_variational_2017, rezende_variational_2015, kingma_variational_2015, ranganath_black_2014}. Generally, sampling and particle methods require a large number of evaluations of the computational model, which can be prohibitive for expensive models. Most variational methods, on the other hand, need the gradient of the model outputs with respect to the uncertain parameters, which can be impractical for complex, potentially coupled solvers. \\
In order to reduce the number of computationally expensive forward model evaluations, various surrogate modeling approaches have emerged, ranging from polynomial chaos expansion \cite{marzouk_stochastic_2007, marzouk_stochastic_2009}, adaptive sparse grid collocation methods \cite{ma_efficient_2009}, Bayesian neural networks \cite{mackay_bayesian_1995}, to Gaussian process regression \cite{kennedy_bayesian_2001, rasmussen_gaussian_2003, bilionis_solution_2014, willmann_inverse_2022, semler_adaptive_2023}. Nevertheless, in higher stochastic dimensions, surrogate methods often fail to capture the true posterior as the necessary number of forward evaluations increases exponentially with the number of input dimensions in order to cover the whole input space sufficiently. \\
In addition to conventional black box surrogate modeling approaches, there has been an increasing trend within the research community towards physics-constrained surrogate models. These models integrate prior knowledge of the underlying physics, governing equations, or constraints into the modeling process. Due to the additional information also higher dimensional inverse problems or uncertainty quantification tasks can be tackled as demonstrated in \cite{zhu_physics-constrained_2019, rixner_probabilistic_2021, raissi_physics-informed_2019}. However, for the application of these approaches, the surrogate model has to be adjusted to each investigated forward model, which is often inconvenient when, like in our case, quite different classes of problems should be tackled. Hence, our attempt in this paper is to develop and present efficient black box surrogate approaches. \\
\cite{bilionis_solution_2014} shows how the epistemic uncertainty due to a limited number of forward solver evaluations can be incorporated into the posterior of the uncertain parameters using a Bayesian surrogate such as a Gaussian process. Similarly, \cite{zhang_surrogate-based_2020} models the surrogate uncertainties and extends on this idea by choosing the training samples systematically in high posterior probability regions to reduce the necessary number of forward model calls. Nevertheless, for high dimensional and independent model outputs, even the Gaussian process model can become prohibitively expensive as an independent Gaussian process has to be trained and evaluated for each output. In such cases, it can be beneficial to approximate the (unnormalized) log likelihood function instead of the model outputs, as only a scalar value has to be approximated. \cite{kandasamy_bayesian_2015} follows this approach and proposes to select new training samples by maximizing the exponentiated variance of the surrogate model response. Similarly, \cite{wang_adaptive_2018} chooses new training samples by maximizing the entropy of the approximated posterior distribution. However, both approaches do not consider the boundedness of the log likelihood function, which can serve as substantial additional information for the surrogate approximation without the necessity for additional forward solver evaluations. \\
In this work, we approximate the log likelihood function by a Gaussian process to reduce the number of computationally expensive forward model evaluations. Further, we exploit the knowledge about the boundedness of the log likelihood function by constraining the surrogate response for improved predictive accuracy. To avoid overconfident and potentially inaccurate results, we incorporate the epistemic uncertainty of the surrogate due to limited training points into the posterior distribution of the uncertain parameters. The scalability of our approach to higher dimensions is additionally enabled by an adaptive sampling scheme adopted from \cite{kandasamy_bayesian_2015}. \\
The rest of the paper is structured as follows: In section \ref{sec:methodology}, we introduce our novel approach to solve Bayesian inverse problems. In section \ref{sec:examples}, we investigate three numerical examples to test our approach. Finally, we provide a conclusion in section  \ref{sec:conclusion}.
\footnote{Generally, we use plain letters for scalar, boldface letters for vector-valued, and capital letters for matrix-valued quantities. The code for this paper is implemented in the in-house code QUEENS \cite{biehler_queens_2019}.}

\section{Methodology}
\label{sec:methodology}

Consider a computationally expensive forward model $\mathcal{M}: \mathbb{R}^{n_x} \to \mathbb{R}^{n_{\mathrm{obs}}}$ which takes the parameters $\boldsymbol{x} \in \mathbb{R}^{n_x}$ as inputs. For given noisy observations of the model output $\yobs \in \mathbb{R}^{n_{\mathrm{obs}}}$, the goal is to infer the corresponding input parameters $\boldsymbol{x}$. The one-dimensional vector $\yobs$ can consist of measurements at different times, at different locations, of vector-valued quantities, or taken from different fields.
For simplicity, we assume that the measurement error $\boldsymbol{\epsilon}$ is additive:
\begin{align}
    \yobs = \mathcal{M}(\boldsymbol{x}) + \boldsymbol{\epsilon}.
\end{align}
In a Bayesian framework, we aim to find a posterior distribution of the input parameters:
\begin{align}
    p(\boldsymbol{x}|\yobs) \propto p(\yobs|\boldsymbol{x}) p(\boldsymbol{x}),
    \label{eqn_2_generel_posterior}
\end{align}
where $p(\yobs|\boldsymbol{x})$ denotes the likelihood and $p(\boldsymbol{x})$ the prior distribution. Here, we model the likelihood by a normal distribution with covariance $\Sigma_n$: 
\begin{align}
    p(\yobs|\boldsymbol{x}) &\propto \exp\left(-\frac{1}{2} \norm{\yobs-\mathcal{M}(\boldsymbol{x})}^2_{\Sigma_n}\right). 
\end{align}
 This is just one popular choice and the extension of our approach to other distributions that are based on a distance measure between $\yobs$ and  $\mathcal{M}(\boldsymbol{x})$ is straightforward. 

\subsection{Surrogate Modeling of the Log Likelihood}

\begin{figure}[h!]
    \centering
    \includegraphics[width=0.99\linewidth, clip=True, trim={0 0 0 0.5cm}]{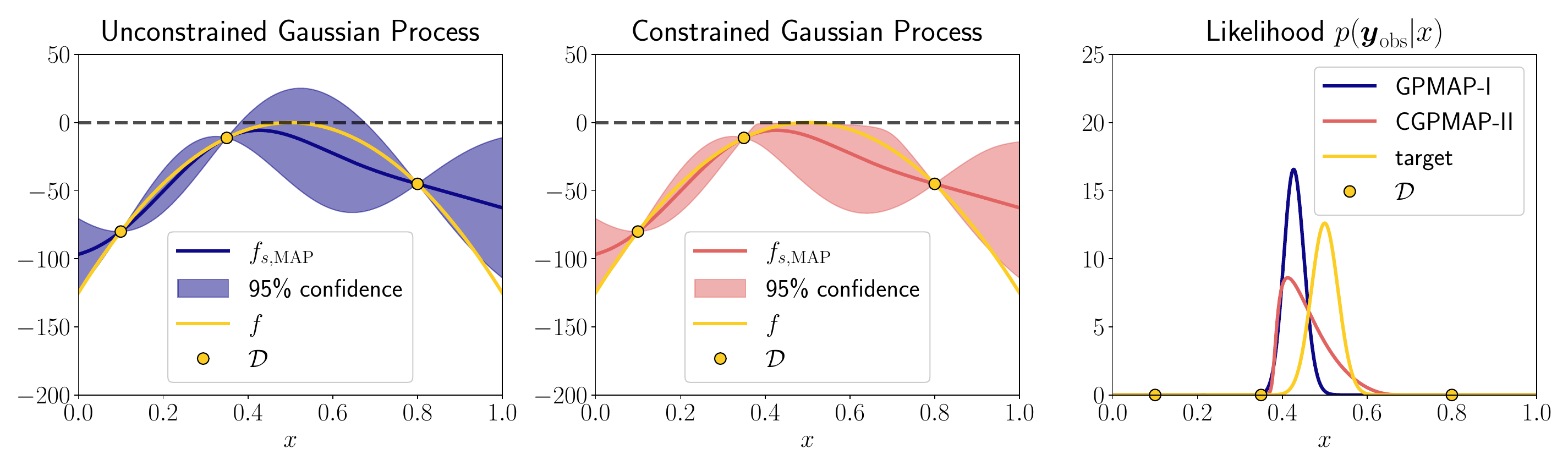}
    \caption{Approximation of the likelihood $p(\yobs|\boldsymbol{x})=\mathcal{N}(0.5, 0.001)$ with three training points. The left plot shows the unconstrained Gaussian process approximation of the log likelihood (up to a constant). The middle plot shows the constrained Gaussian process approximation. The right plot shows the target likelihood and the approximations using the GPMAP-I and CGPMAP-II approach with $q=0.9$.}
    \label{fig:1d_transformation}
\end{figure}

\begin{figure}[h!]
    \centering
    \includegraphics[width=0.7\linewidth, clip=True, trim={0 0 0 0cm}]{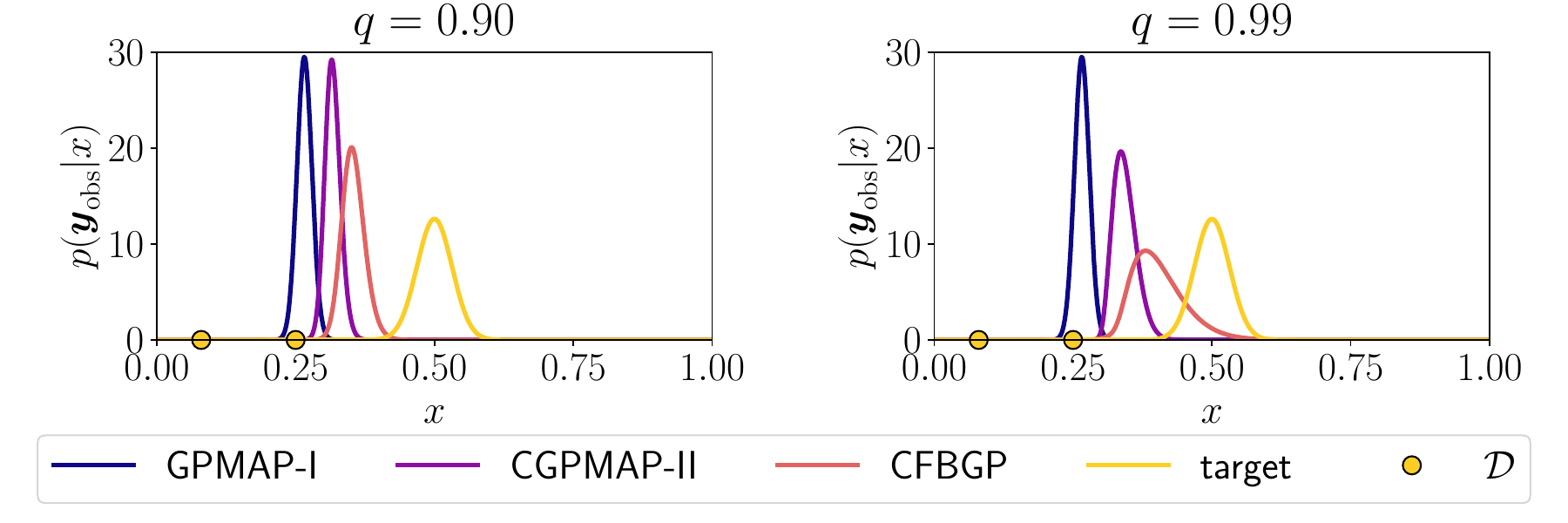}
    \caption{Approximation of the likelihood $p(\yobs|\boldsymbol{x})=\mathcal{N}(0.5, 0.001)$ with two training points. It shows the target likelihood and the approximations using the GPMAP-I, CGPMAP-II, and CFBGP ($n_{\theta}=100$) approach with $q=0.9$ (left) and $q=0.99$ (right).}
    \label{fig:1d}
\end{figure}

For complex forward models, the posterior distribution in \eqref{eqn_2_generel_posterior} is typically not available analytically. Furthermore, classical sampling methods such as Markov Chain Monte Carlo (MCMC) would require an enormous amount of likelihood evaluations, which is infeasible for computationally expensive forward models. Therefore, we propose to approximate the log likelihood (up to a constant) by a surrogate model in order to reduce the number of likelihood evaluations drastically compared to classical sampling methods. To be precise, the exponent of the likelihood function $f(\boldsymbol{x})=-\frac{1}{2} \norm{\yobs-\mathcal{M}(\boldsymbol{x})}^2_{\Sigma_n}$ is approximated by a Gaussian process $f_s \sim \mathcal{GP}(\mu(\boldsymbol{x}), k(\boldsymbol{x}, \boldsymbol{x}'))$, which is defined by a mean function $\mu(\boldsymbol{x})$ and covariance function $k(\boldsymbol{x}, \boldsymbol{x}')$ \cite{rasmussen_gaussian_2006}:
\begin{align}
    \mu(\boldsymbol{x}) &= \mathbb{E}[f_s(\boldsymbol{x})] \nonumber \\
    k(\boldsymbol{x}, \boldsymbol{x}') &= \mathbb{E}[(f_s(\boldsymbol{x})-\mu(\boldsymbol{x}) (f_s(\boldsymbol{x}')-\mu(\boldsymbol{x}')].
\end{align}
The distribution of the function values $\boldsymbol{f}_s$ at $N$ given points $X \in \mathbb{R}^{N \times n_x}$ can be derived as:
\begin{align}
    \boldsymbol{f}_s|X \sim \mathcal{N}(\boldsymbol{\mu}(X), K(X, X)),
\end{align}
where the covariance matrix is evaluated as $K(X,X)_{i,j} = k(X_i, X_j)$. For a given set of training observations $\mathcal{D} = \{(\boldsymbol{x}^{(i)}, f^{(i)})|i=1,...,N)\}$ the function value $f_s$ at a point $\boldsymbol{x}$ follows a normal distribution \cite{rasmussen_gaussian_2006, agrell_gaussian_2019}:
\begin{align}
    f_s | \boldsymbol{x}, \mathcal{D} &\sim \mathcal{N} (m, s^2) \nonumber \\
    m &= \mu(\boldsymbol{x}) +k(\boldsymbol{x}, X) [K(X, X) + \sigma_f^2 I_N]^{-1}(\boldsymbol{f} - \boldsymbol{\mu}(X)), \nonumber \\
    s^2 &= k(\boldsymbol{x}, \boldsymbol{x}) - k(\boldsymbol{x}, X) [K(X, X) + \sigma_f^2 I_N]^{-1} k(X, \boldsymbol{x})
    \label{eqn_2_gp_pred}
\end{align}
where $\sigma_f^2$ denotes an independent and identically distributed Gaussian measurement noise on the observed function values $\boldsymbol{f}$.
Here, we only consider deterministic forward solvers, meaning we can observe the log likelihood function without uncertainty, as we do not consider model inaccuracies. Nevertheless, we optimize the variance term as this can often lead to improved predictive accuracy in sparse data scenarios \cite{gramacy_cases_2012}. As a covariance function, we use the squared exponential kernel with signal variance $\sigma_s^2$ and characteristic lengthscales $\boldsymbol{l} \in \mathbb{R}^{n_x}$:
\begin{align}
    k(\boldsymbol{x}, \boldsymbol{x}') = \sigma_s^2 \exp \left(-\sum_{i=1}^{n_x}\frac{(x_i-x_i')^2}{2 l_i^2}\right),
\end{align}
but there is no restriction to choose another kernel.
Besides the training set $\mathcal{D}$, the surrogate response $f_s$ is also dependent on the hyperparameters $\boldsymbol{\theta} = [\sigma_f^2, \sigma_s^2, \boldsymbol{l}]$. \\ 
Evidently, the approximated unnormalized log likelihood function $f(\boldsymbol{x}) = -\frac{1}{2} \norm{\yobs-\mathcal{M}(\boldsymbol{x})}^2_{\Sigma_n}$ can not be larger than zero as $\norm{\yobs-\mathcal{M}(\boldsymbol{x})}^2_{\Sigma_n} \geq 0$. We can even find a lower estimate $\hat{b} < 0$ for the upper bound of $f(\boldsymbol{x})$, which further improves the predictive accuracy of the surrogate. The estimate is based on the distribution of 
\begin{align}
    b = -\frac{1}{2} \norm{\yobs-\mathcal{M}(\boldsymbol{x})}^2_{\Sigma_n} = -\frac{1}{2} \gamma, \quad \quad \gamma \sim \chi^2(n_{\mathrm{obs}}),
\end{align}
where $\chi^2$ denotes the chi-squared distribution (see \hyperref[appendix:upper_bound]{Appendix}).
We choose a conservative estimate for the upper bound $\hat{b}$ that satisfies $Pr(b \leq \hat{b}) = 0.95$, meaning that the probability that the actual upper bound $b$ is larger than $\hat{b}$ is only $5\%$. In the unlikely event that we encounter a training point with a value larger than $\hat{b}$, we update the upper bound to this value.
The constraint is enforced on the Gaussian process based on virtual observation points as described in \cite{agrell_gaussian_2019}. To enable efficient predictions, the approach is simplified by using only one virtual observation point $\boldsymbol{x}^v$ that coincides with the point $\boldsymbol{x}$, meaning that we do not enforce the constraint in the whole domain, but only at $\boldsymbol{x}^v = \boldsymbol{x}$. Conditioned on the event $\zeta := f_s(\boldsymbol{x}^v) \leq \hat{b} $ this leads to a truncated normal predictive distribution (see \hyperref[appendix:constrained_gp]{Appendix}):
\begin{align}
    p(f_s|\boldsymbol{x}, \zeta, \boldsymbol{\theta}, \mathcal{D}) &= \mathcal{TN}(m, s^2, -\infty, \hat{b})) \nonumber \\ 
    &= \frac{1}{0.5 (1 + \mathrm{erf}(\frac{\hat{b}-m}{\sqrt{2}s}))} \frac{1}{\sqrt{2\pi}s} \exp \left(-\frac{(f_s-m)^2}{2s^2}\right) \mathbb{I}_{f_s \leq \hat{b}},
\end{align}
where the mean $m$ and the variance $s^2$ are defined as in Equation \eqref{eqn_2_gp_pred} and depend on the hyperparameters $\boldsymbol{\theta}$. The posterior of the hyperparameters can be obtained using Bayes' rule:
\begin{align}
    p(\boldsymbol{\theta}|\mathcal{D}) \propto p(\mathcal{D}|\boldsymbol{\theta})p(\boldsymbol{\theta}),
    \label{eqn_2_hyper_post}
\end{align}
with the likelihood $p(\mathcal{D}|\boldsymbol{\theta}) = \mathcal{N}(\boldsymbol{0}, K(X, X) + \sigma_f^2 I_N)$ \cite{rasmussen_gaussian_2006}. For the prior $p(\boldsymbol{\theta})$, we choose independent exponential distributions with small rate parameters, such that the prior is rather uninformative but still has a regularizing effect. The posterior $p(\boldsymbol{\theta}|\mathcal{D})$ is analytically intractable, but we can draw samples from it using sampling methods such as Hamiltonian Monte Carlo \cite{neal_improved_1994, hoffman_no-u-turn_2014}. The computational cost of evaluating the likelihood $p(\mathcal{D}|\boldsymbol{\theta})$ increases cubically with the number of training samples $N$ due to the inversion of the covariance matrix $K(X, X)$. With more elaborate techniques, the computational cost can be reduced to $\mathcal{O}(N^2)$ \cite{wang_exact_2019}. However, this is beyond the scope of this work, as we only consider problems where the evaluation of the forward model for more than several thousand times would be infeasible. Similar to \cite{agrell_gaussian_2019}, we neglect the dependency  of the hyperparameters on the constraint $\zeta$ to reduce the numerical complexity of the surrogate model, i.e., $p(\boldsymbol{\theta}|\mathcal{D}, \zeta) \approx p(\boldsymbol{\theta}|\mathcal{D})$. 
With the approximated posterior $p(\boldsymbol{\theta}|\mathcal{D}, \zeta)$ the hyperparameters can be marginalized to propagate their uncertainty to the posterior of $f_s$:
\begin{align}
    p(f_s|\boldsymbol{x}, \zeta,  \boldsymbol{\theta}, \mathcal{D}) &= \int p(f_s|\boldsymbol{x}, \zeta, \boldsymbol{\theta}, \mathcal{D}) p(\boldsymbol{\theta}| \mathcal{D}, \zeta) \nonumber \\ 
    &\approx \frac{1}{n_{\theta}}\sum_{j=1}^{n_{\theta}} p(f_s|\boldsymbol{x}, \zeta, \boldsymbol{\theta}_j, \mathcal{D}), \quad  \boldsymbol{\theta}_j \sim p(\boldsymbol{\theta}| \mathcal{D}).
\end{align}
This is referred to as \emph{fully Bayesian Gaussian process regression}, whereas the approach of taking a maximum likelihood estimate for the hyperparameters is called \emph{type II maximum likelihood} or ML-II \cite{lalchand_approximate_2020}. Analogously we will refer to the approach using the maximum-a-posteriori (MAP) estimate for the hyperparameters as MAP-II. \\
In the following, we derive an approach to incorporate the surrogate uncertainties into the approximation of the posterior $p(\boldsymbol{x}|\yobs)$ in \eqref{eqn_2_generel_posterior}. 
As we approximate the likelihood in the log space, we have to transform the uncertainties into the original space where we solve the Bayesian inverse problem. Therefore we introduce the random variable $g := \exp(f_s)$, which serves as our approximation of the unnormalized likelihood. Using the rule for the transformation of random variables, the probability density function for $g$ can be derived as (see \hyperref[appendix:transformation]{Appendix}):
\begin{align}
    p(g|\boldsymbol{x}, \zeta, \mathcal{D}) \approx \frac{1}{n_{\theta}} \sum_{j=1}^{n_{\theta}} \frac{2}{g \sqrt{2\pi}s_j \left(1 + \mathrm{erf}\left(\frac{\hat{b}-m_j}{\sqrt{2}s_j}\right)\right)} \exp \left(-\frac{\left(\ln g-m_j\right)^2}{2s_j^2}\right) \mathbb{I}_{0 \leq g \leq \exp(\hat{b})},
\end{align}
where $m_j$ and $s_j^2$ are defined as the mean and the variance in Equation \eqref{eqn_2_gp_pred} for a given hyperparameter sample $\boldsymbol{\theta}_j$. The cumulative density function (CDF) of this distribution is given as:
\begin{align}
    F_g(g | \boldsymbol{x}, \zeta, \mathcal{D}) \approx \frac{1}{n_{\theta}} \sum_{j=1}^{n_{\theta}} \frac{1 + \mathrm{erf}\left(\frac{\ln g - m_j}{\sqrt{2}s_j}\right)}{1 + \mathrm{erf}\left(\frac{\hat{b} - m_j}{\sqrt{2}s_j}\right)}.
    \label{eqn_2_g_cdf}
\end{align}
The CDF provides the probability that the unnormalized likelihood is smaller or equal to a certain value $g$. So we can prescribe $F_g(g | \boldsymbol{x},\zeta,\mathcal{D})=q$, solve the equation for the corresponding value $g$, and use this value as our approximator for the unnormalized likelihood $Z p(\yobs|\boldsymbol{x})$, where $Z$ is a constant. In other words, the inverse CDF $F_g^{-1}(q)=g$ gives us an approximation $g$ of the unnormalized likelihood, which we are certain about with probability $q$, that the true unnormalized likelihood is smaller or equal to this value.
With increasing quantile $q$ the values of the approximation of the unnormalized likelihood grow in regions with high epistemic uncertainties due to limited training points. The idea is here that we prefer to overestimate the likelihood instead of underestimating it in order to reduce bias and explore regions with high epistemic uncertainty due to limited data points in an adaptive sampling setting (see section \ref{sec:ada_sampling}). 
Analogously, the inverse CDF or upper confidence bound is also commonly used in Bayesian optimization as an acquisition function to choose new training points for the Gaussian process \cite{snoek_practical_2012, auer_using_2003, srinivas_gaussian_2010}.  
Although the inverse CDF of \eqref{eqn_2_g_cdf} can not be written in closed form, it can be solved efficiently using iterative methods such as Newton's or Chandrupatla's method \cite{chandrupatla_new_1997}.
As the cost of evaluating our estimator of the likelihood increases linearly with the number of hyperparameter samples $n_{\theta}$, we also consider a more efficient approximator based on the MAP estimate of $\boldsymbol{\theta}$:
\begin{align}
    \thetaMAP = \argmax_{\boldsymbol{\theta}} p(\boldsymbol{\theta}|\mathcal{D}).
\end{align}
If we approximate the posterior of the hyperparameters with $p(\boldsymbol{\theta}|\mathcal{D}) \approx \delta(\boldsymbol{\theta}- \thetaMAP)$ the inverse CDF can be written as:
\begin{align}
    F_g^{-1}(q) \approx \exp\left[\mathrm{erf}^{-1}\left\{q \left(1 + \mathrm{erf}\left(\frac{\hat{b} - m_{\mathrm{MAP}}}{\sqrt{2}s_{\mathrm{MAP}}}\right)\right) - 1\right\} \sqrt{2} s_{\mathrm{MAP}} + m_{\mathrm{MAP}}\right].
    \label{eqn_2_inverse_cdf_map_theta}
\end{align}
Finally, the posterior distribution of the input parameters in \eqref{eqn_2_generel_posterior} is approximated as:
\begin{align}
    p(\boldsymbol{x}|\yobs, \mathcal{D}) \,\Tilde{\propto}\, F_g^{-1}(q) p(\boldsymbol{x}).
    \label{eqn_2_approx_posterior}
\end{align} 
In the following, we denote the approach using the MAP estimate of the unconstrained Gaussian process predictions $f_s$, and therefore not considering any surrogate uncertainties, as GPMAP-I. 
Whereas, if the Gaussian process is constrained, the MAP estimate for the hyperparameters is used, and the uncertainties are incorporated using the inverse CDF in \eqref{eqn_2_inverse_cdf_map_theta} we refer to this approach as CGPMAP-II. Accordingly, if the hyperparameters are marginalized we refer to the constrained fully Bayesian Gaussian process approach as CFBGP. \\
In Figure \ref{fig:1d_transformation}, the approximation of a generic one-dimensional Gaussian likelihood $p(\yobs|\boldsymbol{x})=\mathcal{N}(0.5, 0.001)$ using three training points is shown. The left plot shows the unconstrained Gaussian process approximation of the unnormalized log likelihood $f$. The middle plot shows the constrained Gaussian process approximation, which narrows the confidence intervals and reduces the epistemic uncertainty. The right plot shows the corresponding true target likelihood and the approximations using the GPMAP-I and CGPMAP-II approach with $q=0.9$. It can be seen that due to the consideration of the surrogate uncertainty, the support of the likelihood approximation is wider and covers the whole support of the target likelihood when using the CGPMAP-II approach compared to GPMAP-I. The approximation using GPMAP-I is overconfident and can lead to more biased posterior approximations.
 \\
Figure \ref{fig:1d} visualizes the different approximations of the same likelihood but with only two training points. The approximation is most biased when using the GPMAP-I approach and the least biased when using the CFBGP approach. The bias is also reduced when using a higher quantile $q=0.99$ compared to $q=0.90$ for CGPMAP-II or CFBGP.

\subsection{Sequential Monte Carlo for Bayesian Inference}
Generally, the normalized posterior in \eqref{eqn_2_approx_posterior} can not be solved analytically. Nevertheless, we can obtain a particle representation of the form \cite{bilionis_solution_2014}:
\begin{align}
    p(\boldsymbol{x}|\yobs, \mathcal{D}) \approx \sum_{i=1}^{n_p} w_i \delta(\boldsymbol{x}-\boldsymbol{x}_i)
\end{align} 
using Sequential Monte Carlo (SMC). Several variations of this method have been studied, but the key idea is to start from a particle presentation of the known prior distribution and sequentially move these particles toward the posterior distribution via bridging distributions. Most algorithms consist of an importance sampling step between the bridging distributions, a resampling step based on the weights $w_i$, and rejuvenation steps to mitigate degenerate weights. The rejuvenation steps are usually performed using MCMC. For technical details the interested reader is referred to  \cite{chopin_sequential_2002, chopin_introduction_2020, del_moral_sequential_2006, doucet_sequential_2001}. Here we use the waste-free algorithm of the Python particles library, which uses the samples of the intermediate MCMC steps as particles \cite{dau_waste-free_2021, chopin_introduction_2020}. Nevertheless, also other methods such as variational inference or gradient-based Monte Carlo methods can be used to compute an approximation of the posterior distribution in \eqref{eqn_2_approx_posterior}.

\subsection{Adaptive Sampling Strategy}
\label{sec:ada_sampling}
The accuracy of the approximated posterior distribution of the input parameters in \eqref{eqn_2_generel_posterior} is strongly influenced by the choice of the training points for the surrogate. A common approach is to draw (quasi-)random samples from the prior distribution and use them as training points. In low stochastic dimensions, this can work well. However, the necessary number of training points increases exponentially with the number of input dimensions with this approach. For expensive forward models, this can become infeasible rapidly. Numerous criteria have been proposed for the selection of optimal training points, such as maximizing the mutual information between the chosen points and the points that are not selected \cite{krause_near-optimal_2008} or minimizing the posterior integrated variance \cite{gorodetsky_mercer_2016}. \\
The main difference between the use case here and most other regression problems is that a high accuracy of the surrogate model is only required in regions of $\boldsymbol{x}$ with high posterior probability. \cite{wang_adaptive_2018} uses this insight and proposes to choose a new training point that maximizes the entropy of the posterior distribution approximation of $\boldsymbol{x}$, which is a random variable itself. Similarly, \cite{kandasamy_bayesian_2015} proposes to choose a new training point where the exponentiated variance of the Gaussian process that approximates the log of the joint probability $p(\boldsymbol{x}, \yobs)$, is maximal. \\
Here we adapt the approach proposed in \cite{zhang_surrogate-based_2020} as shown in Algorithm \ref{alg:adaptive_sampling}: We first draw a few initial samples from the prior distribution to train the surrogate. Here, this means drawing samples from the hyperparameter posterior $p(\boldsymbol{\theta}|\mathcal{D})$ or calculating $\thetaMAP$. Based on this surrogate, we can draw samples from the approximated posterior distribution in \eqref{eqn_2_approx_posterior} and use a few of these samples as additional training samples. 
\begin{algo}[b!]
\centering
\fbox{
\begin{minipage}{.9\linewidth}
    \begin{algorithm}[H]
        \LinesNumbered
        \SetAlgoLined
        \begin{enumerate}
            \item Set $j=0$
            \item Draw samples from prior distribution and evaluate forward model to obtain training set $\mathcal{D}_j = \{(\boldsymbol{x}_i, f(\boldsymbol{x}_i)\}_{i=1}^{N_{0}}, \;  \boldsymbol{x}_i \sim p(\boldsymbol{x})$
            \item Train surrogate model based on $\mathcal{D}_j$
            \item Use SMC to obtain particle representation of intermediate posterior distribution $p(\boldsymbol{x}|\yobs, \mathcal{D}_j) \approx \sum_{i=1}^{n_p} w_i \delta(\boldsymbol{x}-\boldsymbol{x}_i)$
            \item \textbf{do} \begin{enumerate}[leftmargin=1.0cm]
                \item[i.] Set $j=j+1$ 
                \item[ii.]  Draw samples from intermediate posterior distribution and evaluate forward model to obtain training set $\hat{\mathcal{D}}_j = \{(\boldsymbol{x}_i, f(\boldsymbol{x}_i)\}_{i=1}^{N_{\mathrm{ada}}}, \;  \boldsymbol{x}_i \sim p(\boldsymbol{x}|\yobs, \mathcal{D}_{j-1})$
                \item[iii.]  Train surrogate model based on $\mathcal{D}_j = \{\hat{\mathcal{D}}_j, \mathcal{D}_{j-1}\}$
                \item[iv.] Use SMC to obtain particle representation of intermediate posterior distribution $p(\boldsymbol{x}|\yobs, \mathcal{D}_j) \approx \sum_{i=1}^{n_p} w_i \delta(\boldsymbol{x}-\boldsymbol{x}_i)$
            \end{enumerate}
            \textbf{while} $\hat{D}_{\mathrm{CS}}\left(p(\boldsymbol{x}|\yobs , \mathcal{D}_{j}), p(\boldsymbol{x}|\yobs, \mathcal{D}_{j-1})\right) > \alpha_{\mathrm{tol}}$
        \end{enumerate}
    \end{algorithm}
\end{minipage}}
\caption[]{Adaptive sampling strategy. Adapted from \cite{zhang_surrogate-based_2020}.}
\label{alg:adaptive_sampling}
\end{algo} 
Again the surrogate is constructed using all evaluated samples of the forward model to draw samples from the approximated posterior distribution. 
\begin{figure}[b!]
    \centering
    \includegraphics[width=0.71\linewidth, trim={0cm 0.15cm 0 0.22cm}, clip=True]{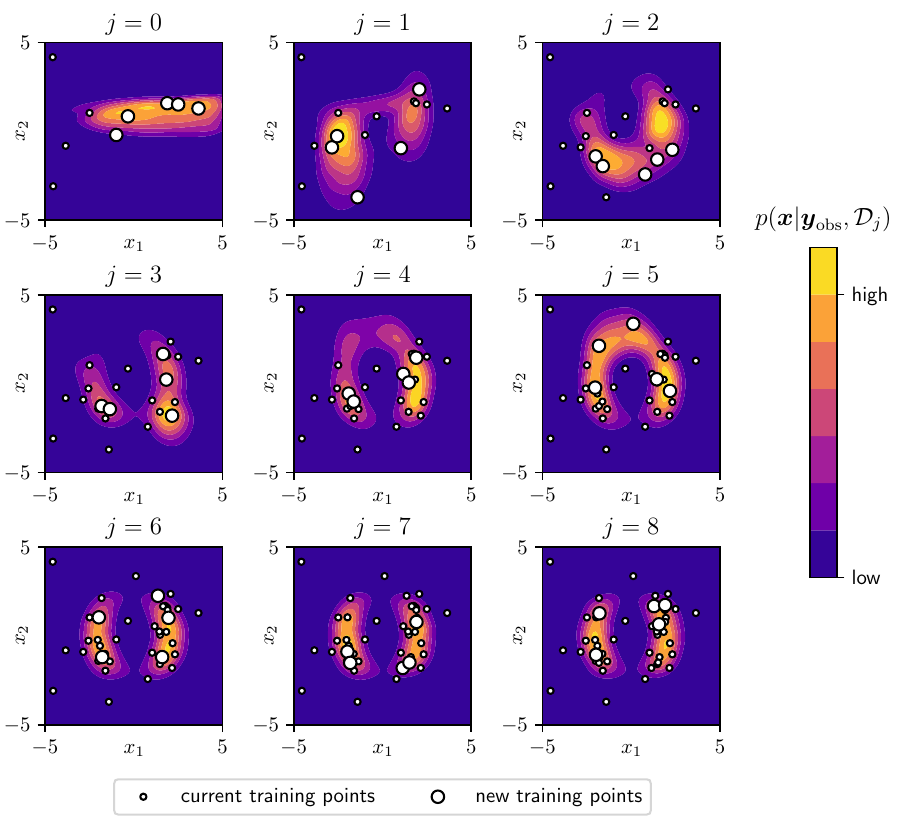}
    \caption{Visualization of adaptive sampling strategy using CFBGP with $n_{\theta}=100$, $N_{0}=5$, $N_{\mathrm{ada}}=5$. Current training points $\mathcal{D}_j$ and new training points $\hat{\mathcal{D}}_j$ are shown for each iteration $j$. The prior distribution $p(\boldsymbol{x})$ is uniform with bounds $[-5,5]$ and the likelihood is taken from \cite{rezende_variational_2015}: $p(\yobs|\boldsymbol{x}) \propto \exp(-U(\boldsymbol{x}))$, where $U(\boldsymbol{x}) = \frac{1}{2}\left(\frac{\norm{\boldsymbol{x}}-2}{0.4} \right)^2 - \ln \left( \exp\left\{-\frac{1}{2}\left[\frac{x_1-2}{0.6}\right]^2\right\} + \exp\left\{-\frac{1}{2}\left[\frac{x_1+2}{0.6}\right]^2\right\} \right)$.}
    \label{fig:banana}
\end{figure}
This process is repeated until the difference between the approximated posterior distributions between the iterations is negligible. \\
To assess the difference between the particle representation of two distributions, we introduce a discrepancy measure that is reliable, computationally cheap, and symmetric. First, we compute a kernel density estimate (KDE) of the particle representation of the distribution with a Gaussian kernel. Subsequently, we can calculate the Cauchy-Schwarz divergence between the two KDEs in closed form \cite{kampa_closed-form_2011}. Still, for a large number of mixture components, meaning particles, this becomes computationally expensive, and in high dimensions, a large number of particles is necessary for an accurate representation of the distribution. Hence, we only compute the divergence measure between the marginal distributions and just use a subset of 5000 particles, which is sufficient for an accurate representation of the distribution in one dimension. In essence, we evaluate the maximal Cauchy-Schwarz divergence between the marginal distributions of the two compared distributions:
\begin{align}
    \hat{D}_{\mathrm{CS}}(p(\boldsymbol{x}), q(\boldsymbol{x})) := \max_{x_j \in \{1, ...,  n_x\}} D_{\mathrm{CS}}(p(x_j), q(x_j)) = \max_{x_j \in \{1, ...,  n_x\}} - \ln \left( \frac{\int p(x_j) q(x_j) dx_j}{\sqrt{\int p(x_j)^2 dx_j \int q(x_j)^2 dx_j} }\right),
    \label{eqn_2:cs-divergence}
\end{align}
using KDE approximations of the particle representations.
In the following, we use the discrepancy measure defined in \eqref{eqn_2:cs-divergence} as a convergence measure by evaluating the measure between intermediate posterior distributions (see Algorithm \ref{alg:adaptive_sampling}). Further, we use the measure to assess the accuracy of posterior approximations by comparing them to a reference posterior approximation. \\
The adaptive sampling strategy using CFBGP with $n_{\theta}=100$, $N_{0}=5$, $N_{\mathrm{ada}}=5$ is visualized in Figure \ref{fig:banana}. The current training points $\mathcal{D}_j$ and newly drawn training points $\hat{\mathcal{D}}_j$ are indicated by white dots for each iteration $j$. For the underlying problem the prior distribution $p(\boldsymbol{x})$ is uniform with bounds $[-5,5]$ and the likelihood is taken from \cite{rezende_variational_2015}: $p(\yobs|\boldsymbol{x}) \propto \exp(-U(\boldsymbol{x}))$, where $U(\boldsymbol{x}) = \frac{1}{2}\left(\frac{\norm{\boldsymbol{x}}-2}{0.4} \right)^2 - \ln \left( \exp\left\{-\frac{1}{2}\left[\frac{x_1-2}{0.6}\right]^2\right\} + \exp\left\{-\frac{1}{2}\left[\frac{x_1+2}{0.6}\right]^2\right\} \right)$.\\
Using adaptive sampling, the number of forward model evaluations can be reduced drastically compared to naively drawing samples from the prior distributions. Nevertheless, there is a chance that not all regions of the true posterior distribution with significant probability are discovered. This risk can be reduced by choosing a high quantile $q$ in \eqref{eqn_2_g_cdf} to weigh the surrogate uncertainties stronger. This underlines the importance of modeling the uncertainties of the surrogate model in order to capture the true posterior distribution with a feasible amount of forward solver evaluations. In the adaptive sampling setting, the quantile $q$ can be viewed as a trade-off factor between exploitation and exploration, as high values of $q$ lead to more exploration, and low values of $q$ lead to more exploitative sampling. Additionally, we would like to point out that the used adaptive sampling strategy allows for batch-wise evaluations of the forward model and consequently the possibility of using the available computational resources efficiently.

\section{Numerical Examples}
\label{sec:examples}
In the following, we test our approach on three numerical examples. First, we investigate the convergence of the algorithm to the true solution for a generic Gaussian likelihood. Subsequently, we infer the diffusivity field for a diffusion problem and finally, we look at a practically relevant example with a coupled forward solver.

\subsection{Artificial Inverse Problem with Generic Gaussian Likelihood}

\begin{figure}[h!]
    \centering
    \includegraphics[width=0.64\linewidth, trim={0 0 0 0cm}, clip=True]{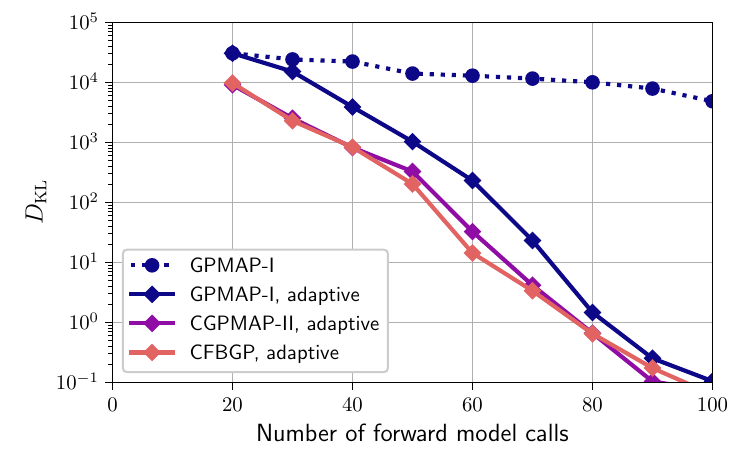}
    \caption{KL-divergence over the number of solver calls for $n_x=10$. The number of initially and adaptively drawn training samples is set to $N_{0}=20$ and $N_{\mathrm{ada}}=10$.}
    \label{fig:gauss_lik_kl}
\end{figure}

\begin{figure}[h!]
    \centering
    \includegraphics[width=0.7\linewidth, trim={0 0 0 0.2cm}, clip=True]{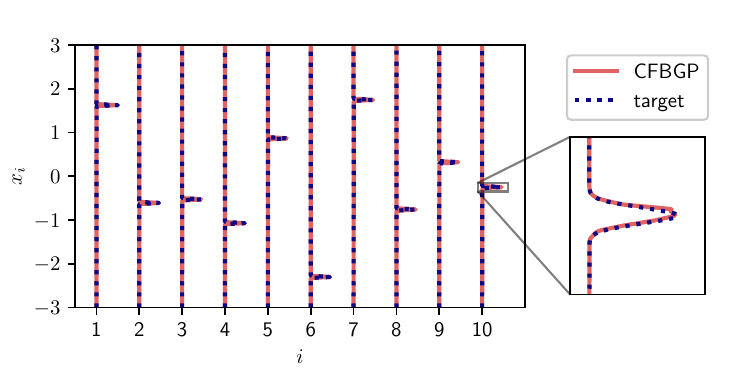}
    \caption{KDE of marginal posterior distributions of the true posterior $p(\boldsymbol{x}|\yobs)$ and an approximation using the CFBGP approach with 100 solver calls and $n_{\theta}=100$.}
    \label{fig:gauss_marginals}
\end{figure}

To test the convergence to the true posterior, we analyze an artificial inverse problem where the posterior distribution of the uncertain parameters can be derived analytically. The prior is given as $p(\boldsymbol{x}) = \mathcal{N}(\boldsymbol{0}, I_{n_x})$ and the likelihood follows a normal distribution $p(\yobs | \boldsymbol{x}) = \mathcal{N}(\boldsymbol{a}, \sigma^2 I_{n_x})$,  where the mean vector is sampled from a standard normal distribution $\boldsymbol{a} \sim \mathcal{N}(\boldsymbol{0}, I_{n_x})$ and the likelihood variance is set to $\sigma^2=10^{-4}$. The posterior distribution can be derived as $p(\boldsymbol{x}|\yobs) = \mathcal{N}(\frac{1}{1+\sigma^{2}}\boldsymbol{a},  \frac{\sigma^2}{\sigma^{2} + 1} I_{n_x})$. We quantify the accuracy of the approximated posterior distribution $p(\boldsymbol{x}|\yobs, \mathcal{D})$ by evaluating the Kullback-Leibler (KL) divergence \cite{kullback_information_1951}:
\begin{align}
    D_{\mathrm{KL}}\big(p(\boldsymbol{x}|\yobs), p(\boldsymbol{x}|\yobs, \mathcal{D})\big) = \int p(\boldsymbol{x}|\yobs) \ln \frac{p(\boldsymbol{x}|\yobs)}{p(\boldsymbol{x}|\yobs, \mathcal{D})} d\boldsymbol{x}.
\end{align}
To evaluate the KL divergence, we approximate the particle representation of $p(\boldsymbol{x}|\yobs, \mathcal{D})$ by a normal distribution, parameterized with the empirical mean and covariance of the particles. 
Subsequently, we can compute the KL divergence between the approximated and the true posterior analytically.
In Figure \ref{fig:gauss_lik_kl}, the KL-divergence is plotted over the number of solver calls for $n_x=10$. For better comparability, the KL-divergence is averaged over ten runs of the algorithm with different training point initializations drawn from the prior. The number of initially and adaptively drawn training samples is set to $N_{0}=20$ and $N_{\mathrm{ada}}=10$. 
We compare the convergence for the three considered approaches GPMAP-I, CGPMAP-II, and CFBGP with quantile $q=0.9$ and $n_{\theta}=100$ hyperparameter samples and use 2000 particles and 25 rejuvenation steps for the SMC algorithm.
It can be seen that when the samples are drawn naively from the prior distribution, the convergence of the approximation to the true posterior distribution is comparatively slow. On the other hand, when the samples are drawn adaptively using Algorithm \ref{alg:adaptive_sampling}, all considered approximations converge to a good approximation of the true posterior within less than 100 solver calls. The marginal posterior distributions of the true posterior and an approximation using the CFBGP approach with 100 solver calls are shown in Figure \ref{fig:gauss_marginals}. The results might imply that the uncertainty of the Gaussian process does not have to be modeled. However, the log likelihood function of the investigated problem is very simple, which is not the case in more realistic applications, such as in the following examples.

\subsection{Inference of Diffusivity Field}
\begin{figure}[h!]
    \centering
    \includegraphics[width=0.82\linewidth, trim={0cm 0cm 0cm 1cm}, clip=True]{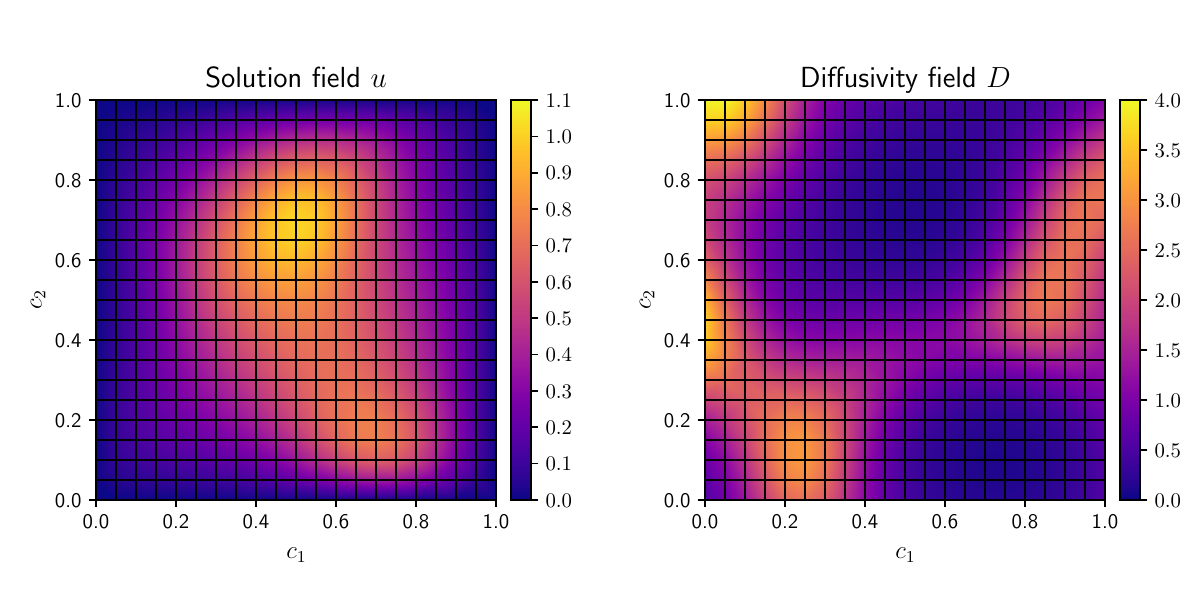}
    \caption{Ground truth of the diffusivity field $D$ and the corresponding solution $u$.}
    \label{fig:ground_truth_diffusion}
\end{figure}

\begin{figure}[h!]
    \centering
    \includegraphics[width=0.82\linewidth, trim={0 0cm 0 1cm}, clip=True]{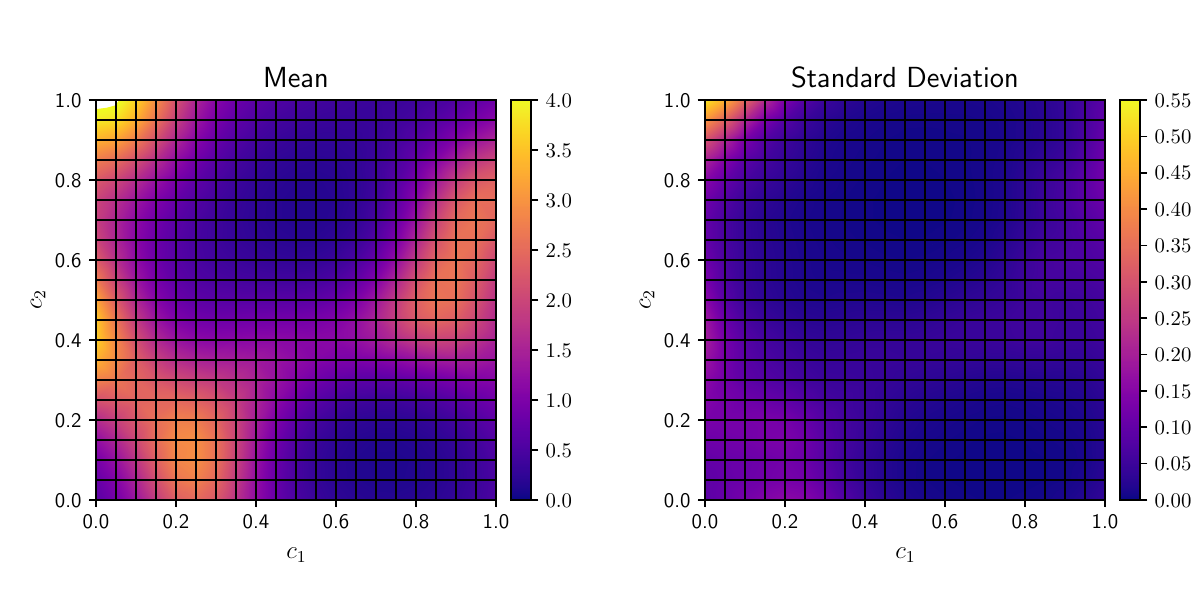}
    \caption{Reference solution of the inverse problem using SMC. The mean and the standard deviation of the posterior of the diffusivity field $D$ are shown.}
    \label{fig:mcmc_diffusion}
\end{figure}

\begin{figure}[h!]
    \centering
    \includegraphics[width=0.92\linewidth, trim={0 0cm 0 0cm}, clip=True]{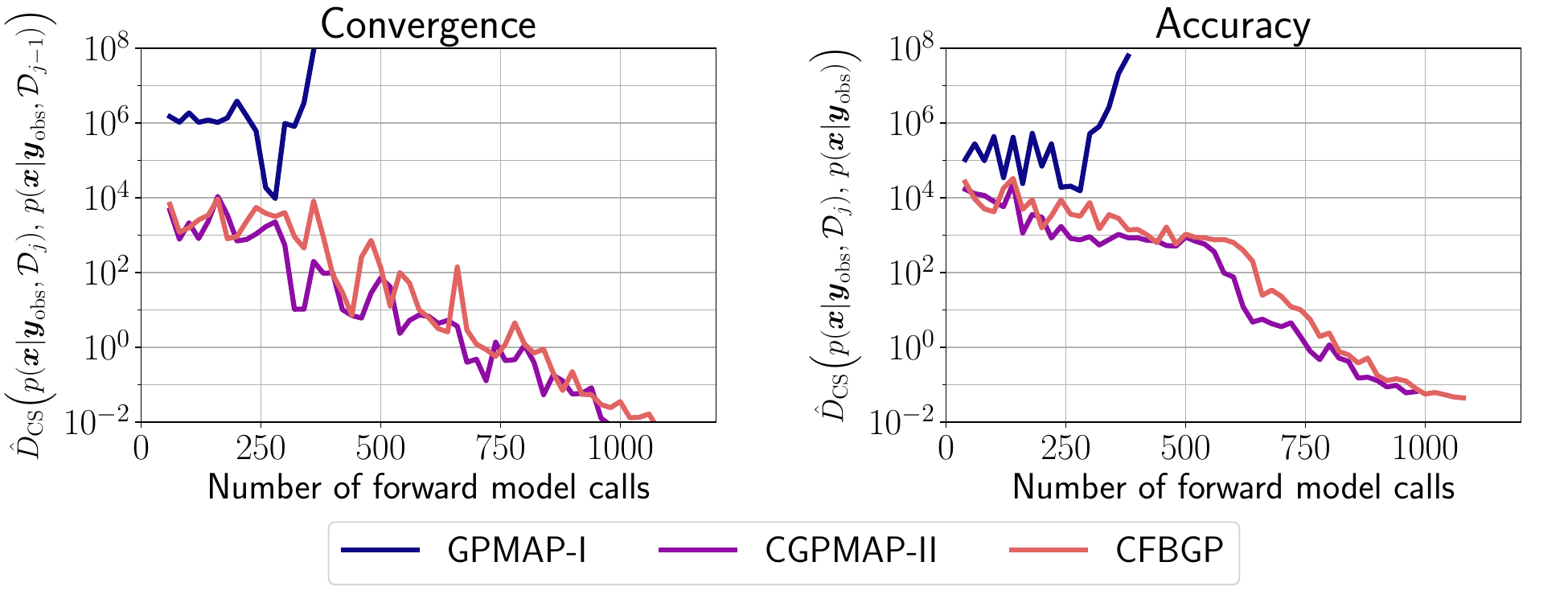}
    \caption{Discrepancy measure defined in \eqref{eqn_2:cs-divergence} between the intermediate posterior distributions $\hat{D}_{\mathrm{CS}}\left(p(\boldsymbol{x}|\yobs, \mathcal{D}_j), p(\boldsymbol{x}|\yobs, \mathcal{D}_{j-1})\right)$ that is used as a convergence measure (left). Discrepancy measure between the posterior approximations and the reference solution using MCMC $\hat{D}_{\mathrm{CS}}\left(p(\boldsymbol{x}|\yobs, \mathcal{D}_j), p(\boldsymbol{x}|\yobs)\right)$, which serves as an accuracy measure (right).}
    \label{fig:cs_div_diffusion}
\end{figure}

\begin{figure}[h!]
    \centering
    \includegraphics[width=0.95\linewidth, trim={0 0.1cm 0 0.6cm}, clip=True]{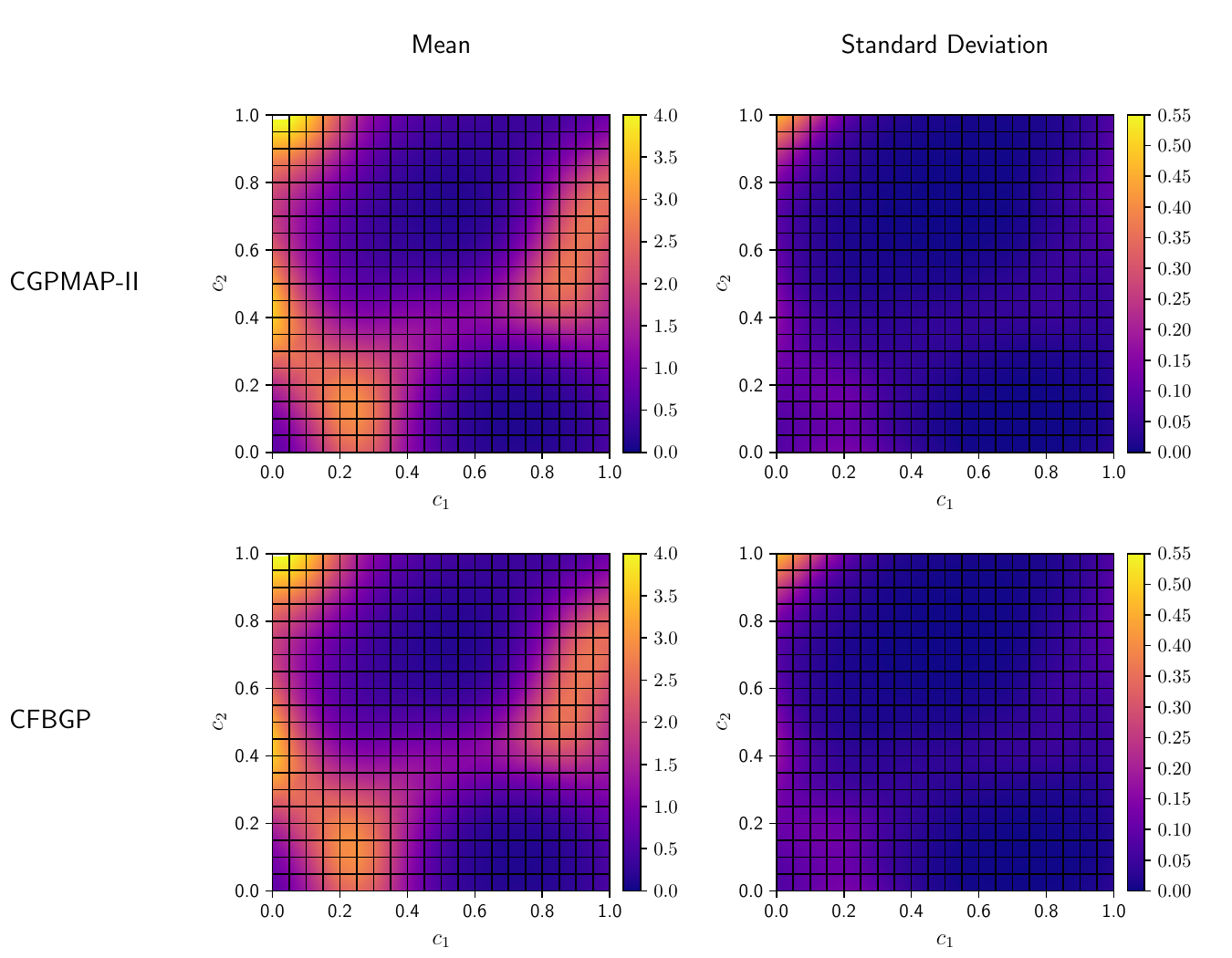}
    \caption{Posterior mean and standard deviation of the diffusivity field $D$ using the CGPMAP-II approach (top row) and using the CFBGP approach with $n_{\theta}=100$ hyperparameter samples (bottom row).}
    \label{fig:posterior_diffusion}
\end{figure}

\begin{figure}[h!]
    \centering
    \includegraphics[width=0.8\linewidth, trim={0 0.1cm 0 0.2cm}, clip=True]{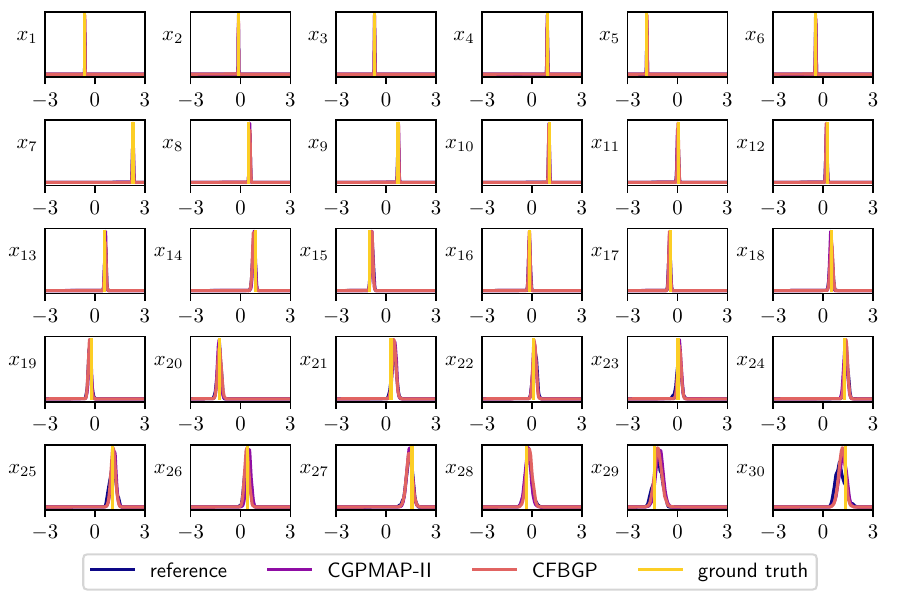}
    \caption{KDE of marginal posterior distributions of the field coefficients $\boldsymbol{x}$ of the reference solution, using the CGPMAP-II approach and using the CFBGP approach with $n_{\theta}=100$ hyperparameter samples and the ground truth.}
    \label{fig:marginals_diffusion}
\end{figure}

In the second example, we consider the following diffusion problem:
\begin{align}
   \nabla \cdot (D(\boldsymbol{c}) \nabla u(\boldsymbol{c})) &= 10, \quad \boldsymbol{c} \in \Omega \\ 
   u(\boldsymbol{c}) &= 0,  \phantom{1} \quad \boldsymbol{c} \in \partial \Omega 
   \label{eqn_3_pde_diffusion}
\end{align}
where $\boldsymbol{c}$ denotes the coordinates in the domain $\Omega$, $D(\boldsymbol{c})$ denotes the diffusivity coefficient at location $\boldsymbol{c}$ and $u(\boldsymbol{c})$ is a spatially varying density. The PDE is solved using the finite element FEniCS library \cite{scroggs_basix_2022, alnaes_unified_2014} with a $20 \times 20$ mesh of bilinear quadrilateral elements.
The prior of the diffusivity coefficient is modeled as a lognormal random field based on a normal random field $G$ with zero mean and covariance function
\begin{align}
   Cov[G(\boldsymbol{c}), G(\boldsymbol{c}')] = \exp \left[ -\frac{\norm{\boldsymbol{c} - \boldsymbol{c}'}_2^2}{2 l_G^2}\right]
   \label{eqn_3_cov_fun}
\end{align}
where $l_G = 0.2$. The prior of the field is discretized using the Karhunen-Loève expansion \cite{ghanem_stochastic_2003}:
\begin{align}
   D(\boldsymbol{c}|\boldsymbol{x}) = \exp\big(G(\boldsymbol{c}|\boldsymbol{x})\big) = \exp\left(\sum_{i=1}^k x_i \phi_i(\boldsymbol{c})\right)
   \label{eqn_3_kle_diffusion}
\end{align}
where $\phi_i$ are the eigenfunctions of the covariance function, weighted by the square root of the corresponding eigenvalues, and the coefficients $x_i$ are normally distributed random variables with unit variance. The number of expansion terms is set to $k=30$, which allows to preserve 99\% of the variance of the field $G$. The goal is to infer a posterior distribution of the coefficients $x_i$ for given observations of the density $u$. In order to create artificial observations, we draw a random sample from the prior distribution $p(\boldsymbol{x})$, reconstruct the diffusivity field using Equation \eqref{eqn_3_kle_diffusion}, solve the discretized forward problem in \eqref{eqn_3_pde_diffusion}, evaluate the density $u$ at all $361$ nodes inside the domain $\Omega$, add independent and normally distributed noise with variance $10^{-4}$ and collect the observations in the vector $\yobs$. The generated diffusivity field and corresponding solution field are shown in Figure \ref{fig:ground_truth_diffusion}. \\
As a reference solution, we solve the inverse problem without a surrogate using SMC with 3000 particles and 30 rejuvenation steps, which leads to almost three million model evaluations. Figure \ref{fig:mcmc_diffusion} shows the mean and the standard deviation of the obtained posterior distribution.  It can be seen that the mean aligns well with the ground truth solution and that the uncertainty of the posterior is maximal in the bottom corners.  \\
For our proposed approach we set $N_{0}=40$ and $N_{\mathrm{ada}}=20$ and use 3000 particles and 30 rejuvenation steps for the SMC algorithm. As a convergence criterion, we choose $\alpha_{\mathrm{tol}}=10^{-2}$. The convergence of the algorithm using the CGPMAP-II approach and using the CFBGP approach with $q=0.90$ and $n_{\theta}=100$ hyperparameter samples is visualized in Figure \ref{fig:cs_div_diffusion}. It shows the discrepancy measure defined in \eqref{eqn_2:cs-divergence} between the intermediate posterior distributions $\hat{D}_{\mathrm{CS}}\left(p(\boldsymbol{x}|\yobs, \mathcal{D}_j), p(\boldsymbol{x}|\yobs, \mathcal{D}_{j-1})\right)$ that is used as a convergence measure.
Further, it shows the discrepancy measure between the posterior approximations and the reference solution using MCMC $\hat{D}_{\mathrm{CS}}\left(p(\boldsymbol{x}|\yobs, \mathcal{D}_j), p(\boldsymbol{x}|\yobs)\right)$, which serves as an accuracy measure. \\
The convergence criterion is met after 980 solver calls using the CGPMAP-II approach and after 1080 solver calls if the CFBGP approach is used. However, the discrepancy $\hat{D}_{\mathrm{CS}}$ to the reference solution indicates that the accuracy of the approximated posterior is slightly higher when using the CFBGP approach. 
Nevertheless, the approximated posterior mean and standard deviation of the field are very accurate in both cases, as they align well with the reference solution (see Figure \ref{fig:mcmc_diffusion} and \ref{fig:posterior_diffusion}). The high accuracy is also evident when looking at the posterior marginals of the inferred coefficients in Figure \ref{fig:marginals_diffusion}. It can be seen that the marginal posterior approximations align well with the reference solution. As the coefficients $x_i$ correspond to the weighted eigenfunctions $\phi_i$ with decreasing eigenvalue with increasing index $i$, the marginal posterior variance of $x_i$ increases with increasing index $i$. It must be noted that due to the added measurement noise and the used prior information, the ground truth does not usually align with the mode of the posterior marginals but lies within regions of significant probability mass. \\
On the other hand, when using the GPMAP-I approach, the SMC algorithm becomes unstable and terminates due to numerical issues after 320 solver calls. Due to the exploitative nature of the GPMAP-I approach, the algorithm puts new training points in local maxima of the posterior far off the global maximum, and due to the unboundedness and exponentiation of the Gaussian process, the posterior approximations become very sharp with small variance. Ultimately, this leads to a failed Cholesky decomposition in the MCMC step of the SMC algorithm due to extremely close training points. This highlights the importance of considering the surrogate uncertainties for more complex likelihoods if the adaptive sampling strategy in Algorithm \ref{alg:adaptive_sampling} is applied, as this encourages exploration and prevents getting stuck in local optima.
 \\

\subsection{Inference of Stiffness Field in a Human Lung Model}
\begin{figure}[h!]
    \centering
    \includegraphics[width=0.9\linewidth, trim={0 0cm 0 0cm}, clip=True]{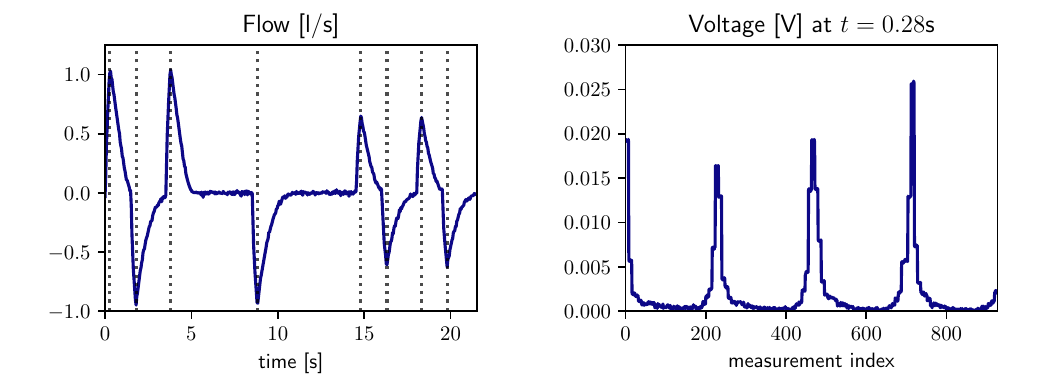}
    \caption{Flow measurements at the ventilator (left). The dotted lines indicate the eight points in time when the EIT measurements are conducted. The right plot shows the measurements of all 32 electrodes at the first time of measurement. In total, there are 32 configurations with 29 measurements leading to 928 measurements at each time step, which are indicated here by the measurement index.}
    \label{fig:measurements_lungs}
\end{figure}

\begin{figure}[b!]
    \centering
    \includegraphics[width=0.99\linewidth, trim={3mm 0cm 3mm 0cm}, clip=True]{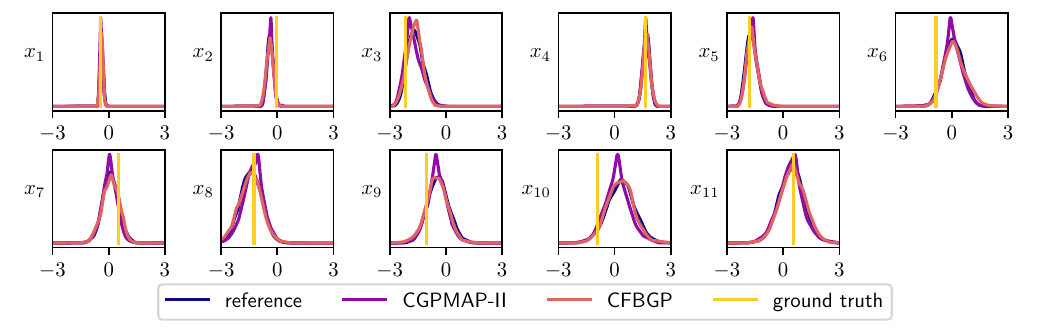}
    \caption{KDE of marginal posterior distributions of the field coefficients $\boldsymbol{x}$ of the reference solution, using the CGPMAP-II approach, using the CFBGP approach with $n_{\theta}=100$ hyperparameter samples and the ground truth.}
    \label{fig:marginals_lungs}
\end{figure}

To this day, respiratory diseases are one of the leading causes of death and affected patients often require mechanical ventilation as a life-saving measure. Especially for diseased, i.e., heterogeneous lungs, adequate adjustment of ventilation parameters is a balancing act between sufficient oxygenation of the blood and removal of carbon dioxide from the lungs, and minimizing damage to lung tissue, a potential side effect triggered and exacerbated by ventilation. Due to the high inter-patient variability, a patient-specific ventilation strategy is crucial, but also challenging: at the bedside, physicians usually only have the global ventilation measurements at hand to customize the ventilation parameters, which do not deliver insight into the local lung behavior. However, it is the individual heterogeneity of diseased lungs that requires deeper local insight and understanding. Computational and patient-specific models can help provide more information about local phenomena and tissue conditions and ultimately advance the development of protective and patient-specific ventilation strategies. \\
With the presented method to solve inverse problems, we investigate in a simplified application-oriented example a well-established reduced dimensional model of the human lungs. Assuming measurements observable at the bedside (ventilation measurements and electrical impedance tomography, see below), we determine with the presented approach inversely the material parameters of the lung model, which eventually may allow to draw conclusions about the local conditions in a patient's lungs. The lung model was developed in our group (see e.g. \cite{ismail_coupled_2013, roth_coupling_2017}) and is implemented in the in-house code BACI \cite{noauthor_baci_nodate}. The human lungs are supplied with air by a conducting airway tree that begins at the trachea and gradually branches to distal regions of the organ.
In each so-called generation, an airway branches into two smaller airways. In total, the human lungs comprise around 23 generations, with the first 17 generations acting as a purely conducting airway network. Under various assumptions, the Navier-Stokes equations that describe the 3D fully resolved air flow in conducting airways and their visco-elastic wall mechanics can be integrated along the coordinates of the idealized cylindrical airways. This leads to the following simplified equations for each airway element of the model \cite{ismail_coupled_2013, roth_coupling_2017, geitner_approach_2023}:
\begin{flalign}
	C \dfrac{\mathrm{d}}{\mathrm{d}t}\left( \dfrac{1}{2} \left( P_{\mathrm{in}} + P_{\mathrm{out}} \right) - \widetilde{P}_{\mathrm{ext}}\right)  + Q_{\mathrm{out}} - Q_{\mathrm{in}} + C\cdot R_{\mathrm{visc}} \dfrac{\mathrm{d}}{\mathrm{d}t} \left( Q_{\mathrm{out}} - Q_{\mathrm{in}}\right) = 0,  \nonumber \\ 
    \dfrac{I}{2} \dfrac{\mathrm{d}}{\mathrm{d}t} \left(Q_{\mathrm{in}} + Q_{\mathrm{out}}\right) + \dfrac{1}{2} \left( R_{\mathrm{\mu}} + R_{\mathrm{conv}}\right) \cdot \textcolor{black}{\left( Q_{\mathrm{in}} + Q_{\mathrm{out}} \right)}  + P_{\mathrm{out}} - P_{\mathrm{in}} = 0, \label{eq:0D_pipe2}
 \end{flalign}
where $C$ describes capacitive, $I$ inductive and $R$ resistive effects of the system. The airways are coupled by the interfacing pressures $P_{\mathrm{in}}, P_{\mathrm{out}}$ and flow rates $Q_{\mathrm{in}}, Q_{\mathrm{out}}$ at the in- and outlets. In this example, the airways are resolved down to the fifth generation, and the remaining generations are modeled by terminal units which are described by a four-element Maxwell model reproducing the nonlinear viscoelastic behavior of the distal lung regions. For such a reduced-dimensional terminal unit, the following relation can be derived from an Ogden-type material law \cite{ogden_large_1972} to mimic its elastic component:
\begin{align}
    P_{\mathrm{a}} - P_{\mathrm{pl/intr}} = \frac{\kappa}{\eta} \left( \frac{V_{\mathrm{a}}}{V_{\mathrm{a,0}}} \right)^{-1} \left( 1 - \left( \frac{V_{\mathrm{a}}}{V_{\mathrm{a,0}}}\right)^{-\eta}\right),
\end{align}
where $P_{\mathrm{a}}$ denotes the internal alveolar pressure, $P_{\mathrm{pl/intr}}$ the external pleural pressure, $V_{\mathrm{a}}$ the current volume of the terminal unit and $V_{\mathrm{a,0}}$ the initial, stress-free volume. $\kappa$ and $\beta$ determine the stiffness and curvature of this pressure-volume relation, respectively. For further details on the lung model, the reader is referred to \cite{ismail_coupled_2013, roth_coupling_2017, geitner_approach_2023}. We simulate the mechanical ventilation of an endotracheally intubated patient and therefore prescribe the inlet pressure at the proximal end of the tube, connected to the ventilator. Besides the measurement of the resulting flow at the ventilator, the physicians use also electrical impedance tomography (EIT) \cite{brown_electrical_2003} at the bedside to get insight into the local ventilation and to adapt the medical therapy accordingly. Therefore, a belt with 32 electrodes is placed on the chest and small voltages are applied. The measured voltages at the remaining electrodes depend on the local conductivity inside the lungs. We can use the strains in the terminal units in a specific time step to evaluate the local resistivity $\rho_{\mathrm{eff}}$ based on the conductivity of the alveolar tissue $\sigma_{\mathrm{alv}}=0.7284 \,\Omega^{-1}\mathrm{m}^{-1}$ and the tortuosity $\Bar{\tau}=1.71$ \cite{roth_correlation_2015, roth_coupling_2017}:
\begin{align}
    \rho_{\mathrm{eff}} = \frac{\Bar{\tau}}{\sigma_{\mathrm{alv}}} \left(5 \frac{V_{\mathrm{a}}}{V_{\mathrm{a,0}}} -4 \right).
\end{align}
As the current volume of the terminal unit $V_{\mathrm{a}}$ is time dependent, we can solve the Laplace equation for a corresponding time step:
\begin{align}
    \nabla \cdot \left(\frac{1}{\rho_{\mathrm{eff}}(\boldsymbol{c})} \nabla u(\boldsymbol{c})\right) = 0, \quad \boldsymbol{c} \in \Omega,
\end{align}
using EIDORS \cite{adler_uses_2006} where $u$ denotes the voltage and $\Omega$ the thorax domain. \\Diseased lungs of patients that undergo mechanical ventilation often exhibit a very heterogeneous state of lung tissue resulting in inhomogeneous straining and thus ventilation of the organ. 
\begin{figure}[b!]
    \centering
    \includegraphics[width=0.55\linewidth, trim={0 1cm 0 1.5cm}, clip=True]{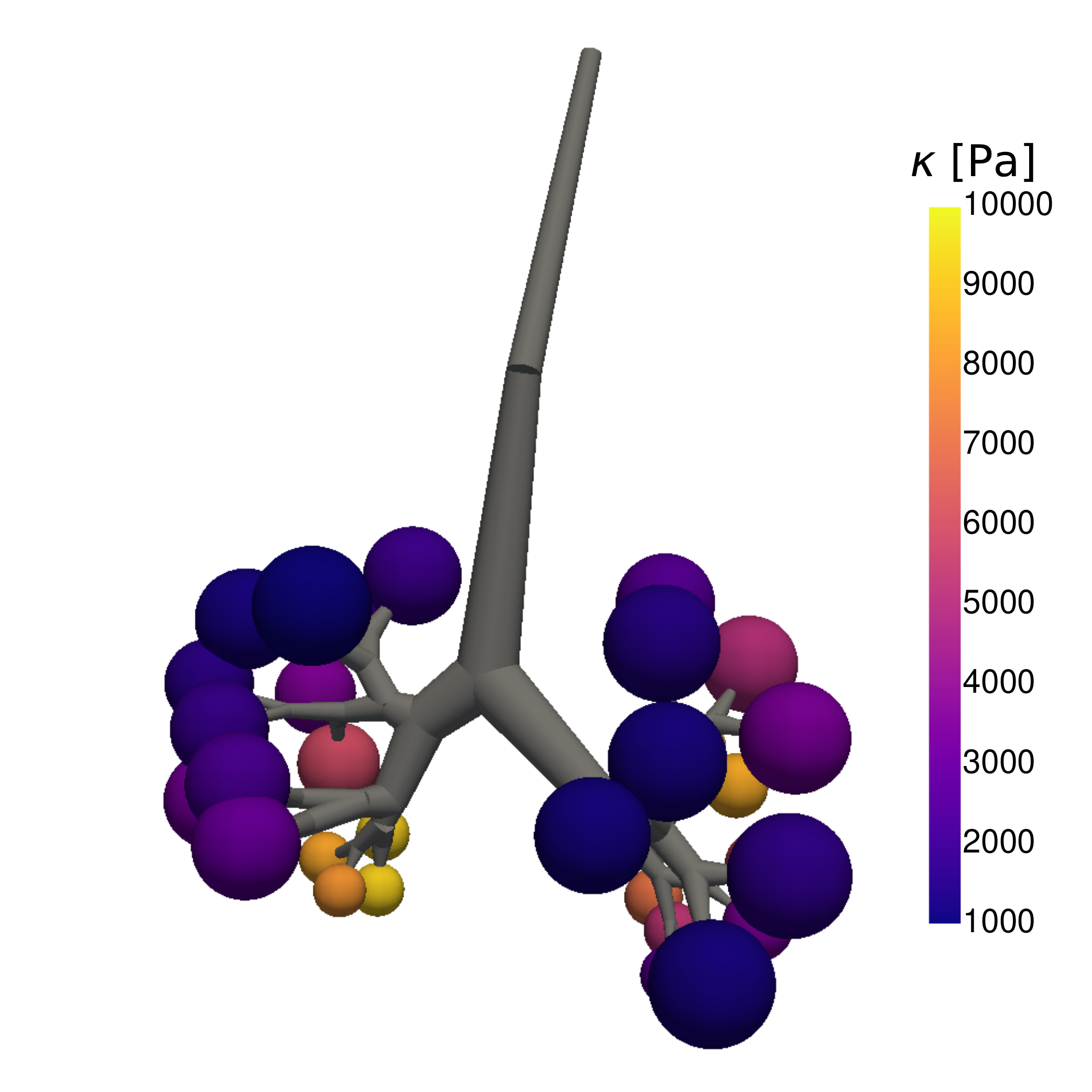}
    \caption{Ground truth of the stiffness field $\kappa$. The airway branches are shown in grey and the colored spheres represent the terminal units.}
    \label{fig:ground_truth_lungs}
\end{figure}
This alternating material behavior is included in the model by a local variation of the stiffness parameter $\kappa$ similar to \cite{roth_coupling_2017}. This functions in the following as quantity of interest in the inverse problem as these are crucial parameters for clinical application and patient-specific models. However, in this study, we will not elaborate further on the clinical case but just use this highly relevant problem as motivation to construct a challenging example. Similarly to the previous numerical example, the prior of the stiffness parameter is modeled as a lognormal random field based on a normal random field $G$ with zero mean and the covariance function in \eqref{eqn_3_cov_fun} with $l_G=100$. Again, the prior of the field is discretized using the Karhunen-Loève expansion:
\begin{align}
   \kappa(\boldsymbol{c}|\boldsymbol{x}) = \kappa_{\mathrm{min}} + \kappa_{\mathrm{scale}} \exp\big(G(\boldsymbol{c}|\boldsymbol{x})\big) = 500 [\mathrm{Pa}] + 2000 [\mathrm{Pa}] \exp\left(\sum_{i=1}^k x_i \phi_i(\boldsymbol{c})\right).
   \label{eqn_3_kle_lungs}
\end{align}
We set $k=11$ such that $99\%$ of the variance of the field $G$ can be reproduced. The choice for the parameters $\kappa_{min}=500 \, \mathrm{Pa}$ and $\kappa_{max}=2000 \, \mathrm{Pa}$ is based on experiments about the mechanical behavior of lung tissue \cite{birzle_coupled_2019, sicard_aging_2018}.
For the inverse analysis, we use flow measurements at the ventilator at 539 time points as well as 928 EIT voltage measurements at eight time points as observations. In this paper, we focus on demonstrating the suitability of our approach for such kinds of problems. Hence, we refrain from using and explaining in detail, real clinical measurements. Thus, analogously to the previous example, the artificial observations are generated by drawing a random sample from the prior $p(\boldsymbol{x})$, solving the reduced dimensional lung model, solving the Laplace equation at the defined eight time steps (see Figure \ref{fig:measurements_lungs}), and adding independent and normally distributed noise to the observations. 
In total, this yields $7424$ voltage measurements and $539$ flow measurements. We set the noise variance for the voltage measurements to $10^{-8}[\mathrm{V}^2]$ and for the flow measurements to $10^{-4}[(\mathrm{l/s})^2]$. The realization of the stiffness field $\kappa$ that is used to generate the observations is shown in Figure \ref{fig:ground_truth_lungs} and the obtained measurements are visualized in Figure \ref{fig:measurements_lungs}.
\begin{figure}[h!]
    \centering
    \includegraphics[width=0.92\linewidth, trim={0 0cm 0 0cm}, clip=True]{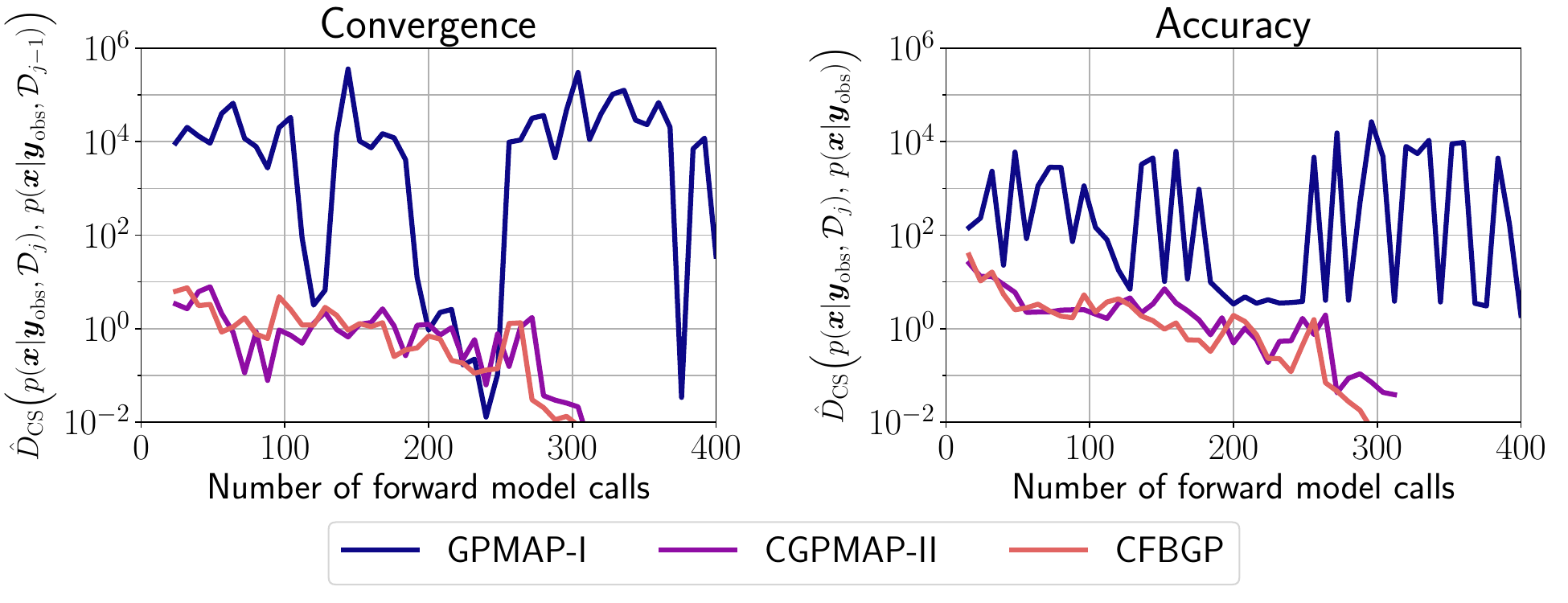}
    \caption{Discrepancy measure defined in \eqref{eqn_2:cs-divergence} between the intermediate posterior distributions $\hat{D}_{\mathrm{CS}}\left(p(\boldsymbol{x}|\yobs, \mathcal{D}_j), p(\boldsymbol{x}|\yobs, \mathcal{D}_{j-1})\right)$ that is used as a convergence measure (left). Discrepancy measure between the posterior approximations and the reference solution using SMC $\hat{D}_{\mathrm{CS}}\left(p(\boldsymbol{x}|\yobs, \mathcal{D}_j), p(\boldsymbol{x}|\yobs)\right)$, which serves as an accuracy measure (right).}
    \label{fig:cs_div_lungs}
\end{figure} \\
We create a reference solution of the posterior $p(\boldsymbol{x}|\yobs)$ without a surrogate model using SMC with 2000 particles and 25 rejuvenation steps, which leads to 652 000 model evaluations.
For our proposed method, we set $N_{0}=16$, $N_{\mathrm{ada}}=8$, $\alpha_{\mathrm{tol}}=10^{-2}$, $q=0.90$ and use 2000 particles and 25 rejuvenation steps for the SMC algorithm. 
Figure \ref{fig:cs_div_lungs} shows that the convergence criterion is met after 312 solver calls using the CGPMAP-II approach and after 304 solver calls using the CFBGP approach with $n_{\theta}=100$. Whereas, when using the GPMAP-I approach, the solution oscillates and does not reach the convergence criterion within 400 model evaluations.  
In this example, the algorithm converges slightly faster when we marginalize the hyperparameters (CFBGP) compared to using the MAP estimate of the hyperparameters (CGPMAP-II). Moreover, the accuracy of the approximated posterior distribution is considerably higher, when using the CFBGP approach. Still, both approaches (CFBGP and CGPMAP-II) deliver a good approximation, which is also visible from the posterior marginals of the field coefficients $\boldsymbol{x}$ in Figure \ref{fig:marginals_lungs}, which shows that the marginal posterior approximations align well with the reference solution using SMC.
\begin{figure}[b!]
    \centering
    \includegraphics[width=0.99\linewidth, trim={3mm 0cm 3mm 0cm}, clip=True]{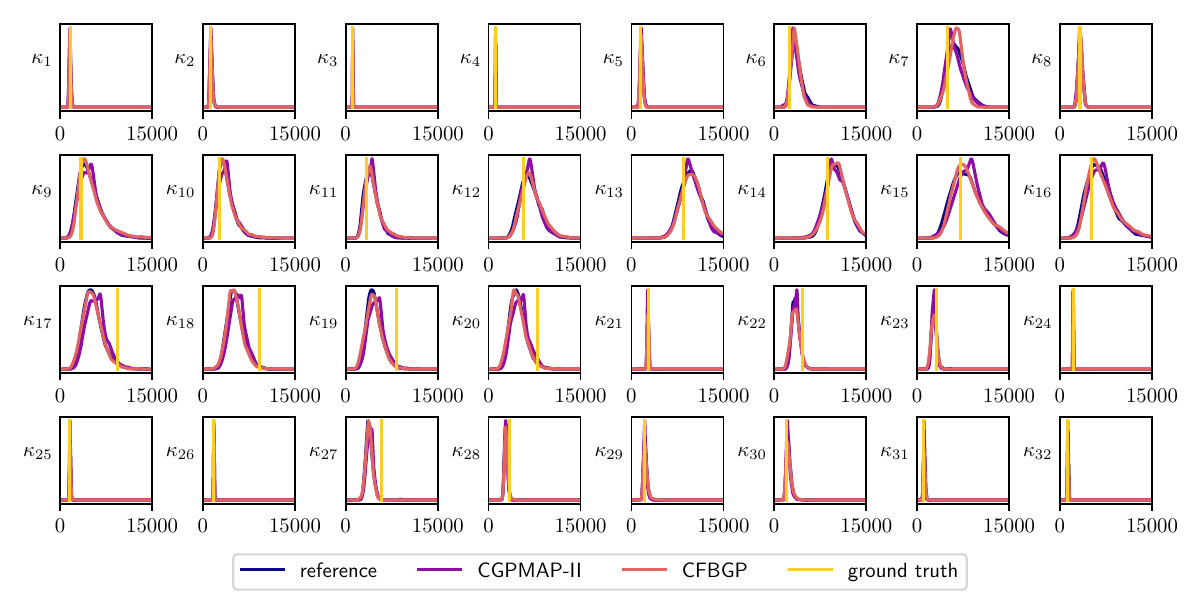}
    \caption{KDE of marginal posterior distributions of the field $\kappa$ at the terminal units (e.g. $\kappa_1 := \kappa(\boldsymbol{c}_1|\boldsymbol{x})$, where $\boldsymbol{c}_1$ is the location of terminal unit $1$) of the reference solution, using the CGPMAP-II approach, using the CFBGP approach with $n_{\theta}=100$ hyperparameter samples and the ground truth.}
    \label{fig:marginals_lungs_field}
\end{figure}
Further, the posterior marginals of the field $\kappa$ at the terminal units in Figure \ref{fig:marginals_lungs_field} support the impression that the approximations align well with the reference solution. Nevertheless, the marginal posterior distributions in Figure \ref{fig:marginals_lungs} also reveal that the approximation using CFBGP is slightly more accurate compared to CGPMAP-II and that the posterior approximation using CGPMAP-II is a little overconfident due to the neglection of the hyperparameter uncertainty. As in the previous example, the ground truth does not usually align with the mode of the posterior marginals but lies within regions of significant probability mass due to the added measurement noise and the used prior information. There is a trend of increasing marginal posterior variance of the coefficients $x_i$ with increasing index $i$, as they correspond to the weighted eigenfunctions $\phi_i$ with decreasing eigenvalue with increasing index $i$. However, the variance of the marginal posteriors of $x_i$ is not strictly increasing with index $i$ due to locally varying influence on the observations, especially to the EIT observations. Generally, the variance of the posterior marginals of the field $\kappa$ at the terminal units increases with increasing distance to the plane of the EIT electrode belt due to the decreasing influence on the measured voltages.

\section{Conclusion}
\label{sec:conclusion}
In this work, we propose an approach to solve Bayesian inverse problems with expensive likelihood models and high dimensional model outputs using Gaussian processes. In this novel approach, the number of likelihood evaluations is drastically reduced by choosing the training samples of the Gaussian process adaptively by sampling from intermediate posterior distributions. In addition, we incorporate the prior knowledge about the boundedness of the log likelihood function in our regression model for increased accuracy of the predictions. As a further improvement, we incorporate the epistemic uncertainty of the surrogate model due to limited training points, where we use the inverse CDF of the transformed predictive distribution. We show accuracy and efficiency in terms of the number of forward model evaluations of the method for a generic problem with analytical solution, as well as for a diffusion problem with 30 uncertain parameters. In a practically relevant example, the stiffness field of a reduced dimensional human lung model that is coupled with an EIT simulation is inferred using only a small number of model evaluations. The given examples, investigating the proposed method in the context of a diffusion problem and a human lung model, underlined the necessity to model the surrogate model uncertainties in high dimensional and complex inverse problems, as the adaptive algorithm does not converge within the given computational bounds when the uncertainty and boundedness of the Gaussian process were neglected (GPMAP-I). Integrating out the hyperparameters increases the computational cost of the Gaussian process predictions significantly, such that only a small number of hyperparameter samples can be used to solve the integral. Therefore, significant improvements are only visible in the final and most complex example when marginalizing the hyperparameters (CFBGP) compared to using the MAP estimate of the hyperparameters (CGPMAP-II). \\
In this paper, we use SMC as inference scheme and Gaussian processes as surrogate models. It must be noted that for the proposed approach, also other inference schemes besides SMC can be used, such as variational inference. In addition, other surrogate models besides Gaussian processes can be employed as long as they possess the ability to quantify the predictive uncertainty and accommodate the upper bound constraint. Here, we only consider Gaussian likelihoods, but the approach can be extended to other likelihoods by modifying the upper bound constraint. 

\section*{Acknowledgments}
The authors gratefully acknowledge financial support by  BREATHE, a Horizon 2020—ERC–2020–ADG project (grant agreement No. 101021526-BREATHE).

\printbibliography

\newpage
\section*{Appendix}
\label{appendix:upper_bound}
\subsection*{Upper bound estimate}
The random variable $\norm{\yobs-\mathcal{M} (\boldsymbol{x})}^2_{\Sigma_n}$ follows a chi-squared distribution:
\begin{align}
    b &= -\frac{1}{2} \norm{\yobs-\mathcal{M}(\boldsymbol{x})}^2_{\Sigma_n}, &\yobs \sim \mathcal{N}(\mathcal{M}(\boldsymbol{x}), \Sigma_n) \nonumber \\
    &= -\frac{1}{2} \boldsymbol{\epsilon}^{\top} \Sigma_n^{-1} \boldsymbol{\epsilon}, &\boldsymbol{\epsilon} \sim \mathcal{N}(\boldsymbol{0}, \Sigma_n) \nonumber \\
    &= -\frac{1}{2} \sum_{i=1}^{n_{\mathrm{obs}}} \hat{\epsilon}_i^2, &\hat{\epsilon}_i \sim \mathcal{N}(0, 1) \nonumber \\
    &= -\frac{1}{2} \gamma, \quad & \gamma \sim \chi^2(n_{\mathrm{obs}})
\end{align}

\subsection*{Simplified constrained Gaussian process}
\label{appendix:constrained_gp}
Following \cite{agrell_gaussian_2019} the predictive distribution of the constrained Gaussian process with one virtual observation point $\boldsymbol{x}^v=\boldsymbol{x}$  and the event $\zeta := f_s(\boldsymbol{x}^v) \leq \hat{b} $  is given as:
\begin{align}
    f_s | \boldsymbol{x}, \zeta, \mathcal{D} \sim \mathcal{N}\big(\mu(\boldsymbol{x}) + A (\zeta - \mu(\boldsymbol{x}^v)) + B(\boldsymbol{f} - \mu(X)), \Sigma \big),
\end{align}
where
\begin{align}
    \zeta \sim \mathcal{TN}\big(\mu(\boldsymbol{x}^v) + A_1 (\boldsymbol{f} - \mu(X)), B_1, -\infty, \hat{b}\big),
\end{align}
and
\begin{align}
    A_1 &= k(\boldsymbol{x}^v, X) (K(X,X) + \sigma_f^2 I_N)^{-1} \nonumber \\
    A_2 &= k(\boldsymbol{x}, X) (K(X,X) + \sigma_f^2 I_N)^{-1} = A_1 \nonumber \\
    B_1 &= k(\boldsymbol{x}^v, \boldsymbol{x}^v) - A_1 k(X, \boldsymbol{x}^v) \nonumber \\
    B_2 &= k(\boldsymbol{x}, \boldsymbol{x}) - A_2 k(X, \boldsymbol{x}) = B_1 \nonumber \\
    B_3 &= k(\boldsymbol{x}, \boldsymbol{x}^v) - A_2 k(X, \boldsymbol{x}^v) = B_1 \nonumber \\
    A &= B_3 B_1^{-1} = I_N \nonumber \\
    B &= A_2 - A A_1 = 0 \nonumber \\
    \Sigma &= B_2 - A B_3^{\top} = 0 .
\end{align}
Consequently, the predictive distribution simplifies to:
\begin{align}
    f_s | \boldsymbol{x}, \zeta, \mathcal{D} 
    \sim \mathcal{TN}\big(\mu(\boldsymbol{x}) + A_2 (\boldsymbol{f} - \mu(X)), B_2, -\infty, \hat{b}\big).
\end{align}

\subsection*{Transformation of continuous random variable}
\label{appendix:transformation}
\begin{align}
    g &= \exp(f_s) =: T(f_s) \nonumber\\
    p_g(g) &= p_{f_s}\left(T^{-1}(g)\right) \left|\frac{d}{dg} T^{-1}(g)\right|  \nonumber\\
    p_g(g|\boldsymbol{x}, \mathcal{D}) &= p_{f_s}\left(\ln g|\boldsymbol{x}, \mathcal{D}\right) \frac{1}{g} \\
    &= \frac{1}{n_{\theta}} \sum_{j=1}^{n_{\theta}} \frac{2}{g \sqrt{2\pi}s_j \left(1 + \mathrm{erf}\left(\frac{\hat{b}-m_j}{\sqrt{2}s_j}\right)\right)} \exp \left(-\frac{\left(\ln g-m_j\right)^2}{2s_j^2}\right) \mathbb{I}_{0 \leq g \leq \exp(\hat{b})},
\end{align}

\subsection*{Consistency of likelihood estimation}
\label{appendix:consistency}
For an infinite number of training points $N$, the predictive distribution of the Gaussian process in \eqref{eqn_2_gp_pred} becomes a Dirac distribution with exact mean, independent of the hyperparameter sample. Consequently, the addends in \eqref{eqn_2_g_cdf} become identical to each other and for $q \in (0, 1)$ the inverse CDF can be written as:
\begin{align}
    \lim_{\substack{s \to 0 \\ m \to \ln g}} \exp\left[\mathrm{erf}^{-1}\left\{q \left(1 + \mathrm{erf}\left(\frac{\hat{b} - m}{\sqrt{2}s}\right)\right) - 1\right\} \sqrt{2} s + m\right] = g, \nonumber
\end{align}
which shows that our chosen estimator for the unnormalized log likelihood is consistent.

\end{document}